# RSA ALGORITHM WITH A NEW APPROACH ENCRYPTION AND DECRYPTION MESSAGE TEXT BY ASCII

Ahmad Steef[1], M. N. Shamma[2] and A. Alkhatib[3]

[1]Department of Mathematics, Al Baath University- Homs- Syria.
[2]Basic Sciences department- the Mechanical and Electrical Engineering- Damascus University- Syria.
[3]Department of mathematics & Faculty sciences - Al Baath University- Homs- Syria.

## *ABSTRACT*

*In many research works, there has been an orientation to studying and developing many of the applications of public-key cryptography to secure the data while transmitting in the systems, In this paper we present an approach to encrypt and decrypt the message text according to the ASCII(American Standard Code for Information Interchange) and RSA algorithm by converting the message text into binary representation and dividing this representation to bytes(8s of 0s and 1s) and applying a bijective function between the group of those bytes and the group of characters of ASCII and then using this mechanism to be compatible with using RSA algorithm, finally, Java application was built to apply this approach directly.*

## 1. INTRODUCTION

Public-Key Cryptography[1 ][2] is the most common major factor for security and protection of Systems. It is a strong technique to secure the data transmitting within Systems. Public-Key Cryptography uses two roles(encrypting and decrypting functions). Encrypting function encrypts massage and converts it to cipher(gibberish) "C", Decrypting Function is applied to "C" to coming back to original message "M". decryption and encryption functions are inverse each other and use distinct keys(numbers)[1]. In a special case, the RSA Algorithm[1 ][2], the most strong common applications of Public-Key Cryptography which published by Rivest, Shamir and Adelman 1978. It depends on the difficult problem in mathematics which is "factorization problem". This Algorithm was built depending on exponential function E(Encryption Function)has inverse D(Decryption Function), but it is difficult to discovering it, because this depends on the " P = NP Problem[2]" is one of the biggest open problems in both mathematics and computer science. RSA Algorithm is used to encrypt and decrypt text according to represent alphabets (A, B, …….Z) as decimal numbers(from 0 ..to 25)[ 2], but what about if the text message includes numbers and some characters like: &@!*....? what about encryption and decryption text message by RSA according to ASCII(American Standard Code for Information Interchange)[3]. This paper focused on cryptography to secure the message while transmitting in the systems, so, protecting confidentiality and integrity of information. In this paper we will illustrate the use of RSA to encrypt and decrypt message which is to be transmitted from sender to receiver by a new approach related of ASCII. It's easy to convert any character in ASCII into associated number but How can we convert big numbers to the associated text by ASCII?. The answer of all inquiries was included in this paper, and a numerical example applied to illustrating the basic idea in applicable way, finally, Java application was built to apply this approach directly.

## 2. RSA ALGORITHM.

RSA Algorithm[1][2] is the most strong common applications of Public-Key Cryptography which published by Rivest, Shamir and Adelman 1978. It uses two distinct keys(two numbers), public-key which possible to be known to every one and the other is private-key which is secured and not allowed to exchange between the sender and reciver. This algorithm described as following:

1- Choose two distinct large random prime numbers p and q.
2- Compute $n = p * q$,
3- Compute Euler's function of n : $\emptyset(n) = (p-1)*(q-1)$.
4- Choose an integer $e$ such that $1 < e < \emptyset(n)$ and $gcd(e, \emptyset(n)) = 1$.,
5- Compute $d$ such that: $e.d = 1(mod\ \emptyset(n))$ .

The number $e$ is the public – key and $d$ is the private –key. Let M is a message need to be encrypted and get cipher C, so, in the first the message M represented as blocks of numbers every value of block must be less than the number "n" [2]. For any block number $M_i\ ; i \in I$ ,such that: $M_i \in Z_n = \{0,1,2,3, \dots \dots n-1\}$, the Encryption role of this algorithm is :$C_i \equiv {M_i}^e (mod\ n)$. and, the Decryption role is :$M_i \equiv {C_i}^d (mod\ n)$.

**Note:** M divided into blocks as form: $(M_1|M_2| \dots \dots . M_i | \dots . .)$and Cipher C will be as form: $(C_1\ |C_2\ | \dots \dots .\ C_i\ | \dots . .)$.

## 3. ADDRESSED PROBLEM

Suppose we have message text and want to encrypt it by RSA algorithm depending on ASCII, first, we will represent it as numbers by convert every character of the message text into associated number depending on ASCII, then we will apply RSA algorithm on those numbers which presented in, but when we want to coming back to original message, there will be problem because there are characters in ASCII represented as decimal numbers of two digits and the others represented as numbers in a three digits, for example the character "a" is represented as 97 and "v" is represented as 118, and " %" is represented as 37, so, for example the "av%" converted to the number "1189737" but according to the number"1189737" we can't discover which message associated by ASCII even if we forwarded to dividing it's digits because we don't know the right mechanism for dividing!, not only for that, but also we want to represent the text as not big decimal numbers as possible as we can, to be able to apply RSA algorithm on that whole text directly in an efficient way.

In the following section we will illustrate our approach to be able to coming back to the original message and how we will use RSA algorithm with this approach.

## 4. PROPOSED SOLUTION AND ASSOCIATED APPROACH WITH RSA.

We know that ASCII table[3] consists of 256 characters and the representation of them as numbers from 0 to 255, so, suppose the group A is equal to $Z_{255} = \{0,1,2 \ldots \ldots 255\}$, and let's define the function $f$ as form: $f: A \rightarrow \{0,1\}^8$. It's easy to know the number of elements of $\{0,1\}^8$ is $2^8 = 256$. Indeed, $f$ is bijective function and so, it has inverse function. We can represent every dicimal number in A as a binary number by using binary and decimal number systems principles[5], and every number consists of 8 of 0s and 1s.

(we can add some 0s to the left of that representation if the number of digits is not 8 ), the same, we can represent every character as binary representation and every binary consists of 8 of 0s and 1s.

Now, we will reivew our approch as algorithm, so suppose the message M which we want to encrypt it, this algorithm illustrates the encryption role and get Cipher "C"(gibberish) as form:

1- Convert every character of M to binary representation(every character must consists of 8 of 0s and 1s) from left to right.
2- Convert whole representation(S) of M to decimal representation $M_1$ .
3- Apply RSA encrypting function on the number $M_1$ and get number $M_2$ .
4- Convert the number $M_2$ to binary representation(but the number of digits must divided by 8...we can do that by adding zeros to the left of representaion)
5- Divide that representation into bytes(8s of 0s and 1s) from left to right.
6- Convert every byte in step 5 into asochiated character in ASCII from left to right and this creates cipher text C.

To come back the original message "M", the following algorithm illustrates that:

1- Convert every character of C to binary representation(every character must consists of 8 of 0s and 1s) from left to right.
2- Convert whole representation of C to decimal representation $M_2$ .
3- Apply RSA decrypting function on the number $M_2$ and will get $M_1$.
4- Convert the number $M_1$ to binary representation(S)(but the number of digits must divided by 8...we can do that by adding zeros to the left of representaion)
5- Divide that representation into bytes( 8s of 0s and 1s) from left to right.
6- Convert every byte in step 5 into asochiated character in ASCII from left to right and this creates the original message text M.

**REMRK (1):** we can stop at step(4)in encryption algorithm above and get cipher as a dicimal number, but in this case we will begin from step(4)in the decryption algorithm mentioned above.

**REMRK (2):** within applying this approch we have to be attantion in conditions of RSA algorithm wich related of message when represented as dicimal number because of the RSA not applicable if this number isn't less than the modulo n (n=p*q) (see [2] ).

## 5. NUMERICAL EXAMPLE.

Suppose the message M as following:

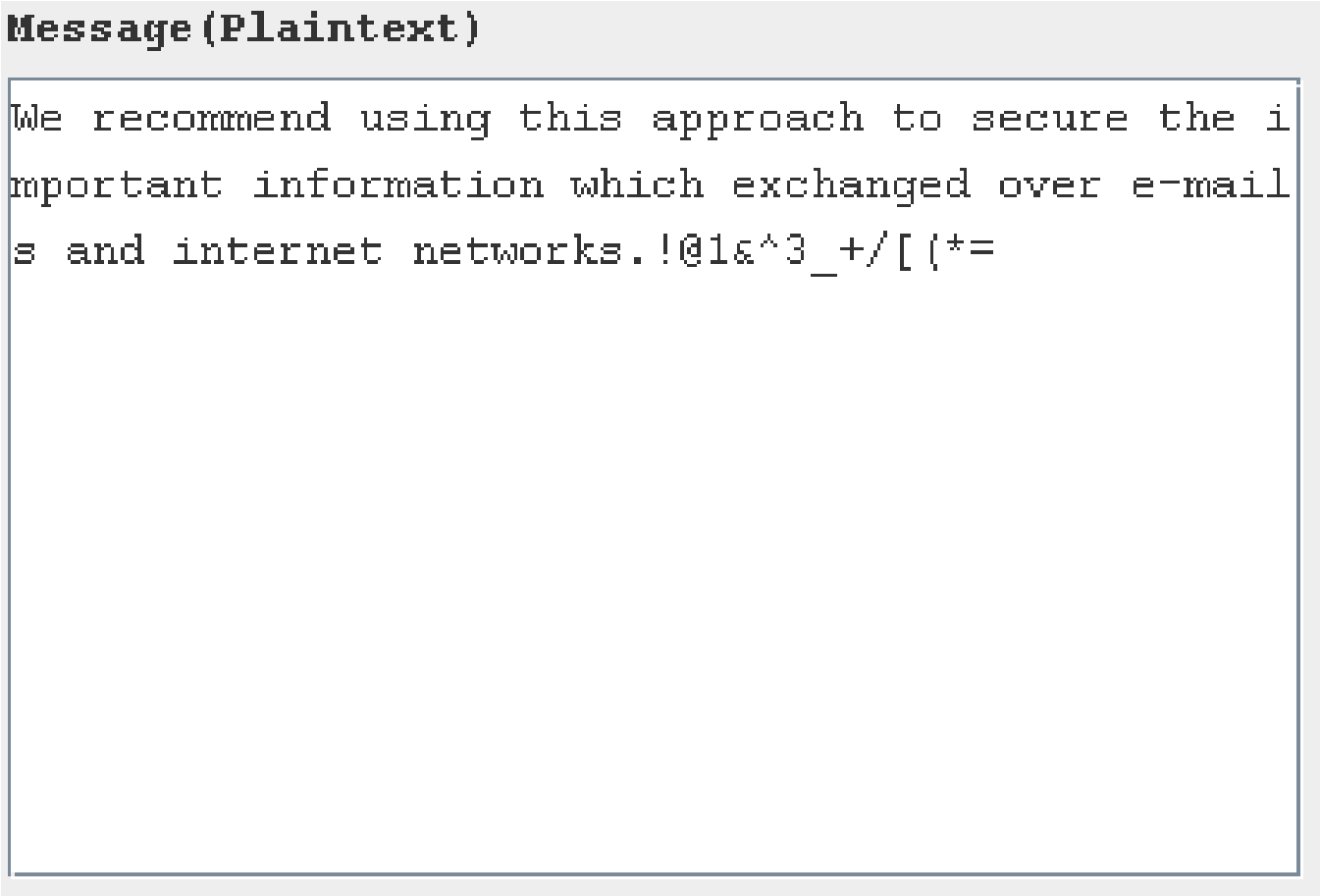

And we want to apply our approach with RSA, so, in the beginning let's build the RSA requirements$(p, q, n\ e, d, \emptyset(n))$ as mentioned above.

By using Miller–Rabin Test algorithm [2][4], we generated a 600-bits prime numbers p and q using java language[4][5] and get the following:

p=36816687633241002182065576654551879280165753023812270675138477824004277853846
17838705799614190168082010883295794014544112269579528986052467638469601985913816
991755581819544236809351.

q=27186794025716014479926827223597630969441403917800649988487626523291402111446
81800611706412181565231784663324402179805861977098492370293532314325853153282910
36135994902620997906 1959.

n=p*q=1000927703394049150956735889190158060942691067859650131675672227138305622
1323278974381732635552684572037879606487115336485429198732097926353032301768089
6095348004285611010290324917226394682382934162876830588655185205717499735657774
2468864641126802282282262224431385457263003407673803603439155820335047816220955 4
367664324531491865426026769622041340025808209957860 9.

So, $\emptyset(n) = (p - 1) * (q - 1) =$
1000927703394049150956735889190158060942691067859650131675672227138305622132327
8974381732635552684572037879606487115336485429198732097926353032301768089609534
8004285611010290324917162391200723977146214838184777035695467892578716161326222
8380083329325482609314349922822027396903406656479729538583915480737541752232029
72453621231991210709688672978694123278837073 00.

Select number e ; $1 < e < \emptyset(n)$, such that: $gcd\big(e, \emptyset(n)\big) = 1$, let take e= 11;

And the number d(inverse product) will be :

d=4549671379063859777076072223591627549739504853907500598525782850628661918783308624718969379796674805471763457494143334766104181241862693796832864440040731606727467527773195014768961926872760180779373401735671683440703086026305280060282856273106042388557315065226921918306349560939347635134266290524912443371601054680566073692363236850440394226304278330856230591.

So, the public- key is (e,n) and private-key is (d, n).Now, let's apply our approach which mentioned above step by step, to encrypt message M.

1- The binary representation of the message (character by character) is: S= 0101011101100101001000000111001001100101011000110110111101101101011010110010101101110011001000010000001110101011100110110100101101110011001110010000001110100011010000110100101110011001000000110000101110000011100000111001001101111011000010110001101101000001000000111010001101111001000000111001101100101011000110111010101110010011001010010000001110100011010000110010100100000011010010110110101110000011011110111001001110100011000010110111001110100001000000110100101101110011001100110111101110010011011010110000101110100011010010110111101101110001000000111011101101000011010010110001101101000001000000110010101111000011000110110100001100001011011100110011101100101011001000010000001101111011101100110010101110010001000000110010100101101011011010110000101101001011011000111001100100000011000010110111001100100001000000110100101101110011101000110010101110010011011100110010101110100011011100110010101110100001000000110111001100101011101000111011101101111011100100110101101110011001011100010000101000000001100010010011001011110001100110101111100101011001011110101101100101000001010100011110 1.

2- The decimal representation of S is: $M_1$=67477990192598109094026434793800128271623759125622767016961491405796088274383644369970131807949813858022588176538096721730314658378073281265611145286262073549835300274884544221576823895398534324930035218598971627656472563528346662621303268839116735117580407292136276840880384768590448944662517840792486693738633291639357.

3- Applying RSA Encryption on $M_1$ and get $M_2$as following:

$$M_2 = M_1^e (mod\ n) =$$

8630628267424397002865353339057570297202073669109555251723582169330998889334187101465152792779967024367063523002136275498607043397142939850644957851666561154751260571623481264893147279833000162802646942400233819241566533597070669145406550404565169899133453077460458404471104791143022106721525602611893089331489685531217659064283488939741765624469536510261947451 3.

4- binary representation of $M_2$ (with 0s which added to the left) is: **10000000**01010001011010000110100010001000000000001100110101110110111010101000011010110001011000000111110110100001110110100010100110100000000110011110010001000110011010011011100111110000101110000100001110111000100000101010111101101010111010110110110101100011000111111101110011101101001001110011101011100110101110010010110010011011011001100111100110110011011110011011011110100010000100010111101111111101100011111011000011001110000111011101011111001011010101011110011001011110100101100001101001001000011101010

1101110011000001011000101111100000010001111101111100011110110110010010111
1111111011110110010001100000111010001101111100011001000111100000100011 1
0110011000110111101110111011011001111000110010001111001111101100101000 11
1110000101001001001011110101000001111111111000001101000111011111110101 00
1000100101111011001111010000000100101101011110001000011100000001010100 0001
0101100110101001100010000010001101111101101000111111111101110101010111 00
1000010101001000110001111000110111111010101010111101101001110101111110 001
0011101111000100111010101000100111011000100110000000111110101000111010 100
0110010001111011001001000010010011000110001111010100111111010110100101 10
01010100110110011011100100100110001110100**1010001**.

5- Divide the representation in step (4) into 8s from "**10000000**" to the last byte which is"**01010001**"
6- Convert every byte from "**10000000**" to "**01010001**" into associated character in ASCII and get the cipher C(more gibberish )as form:

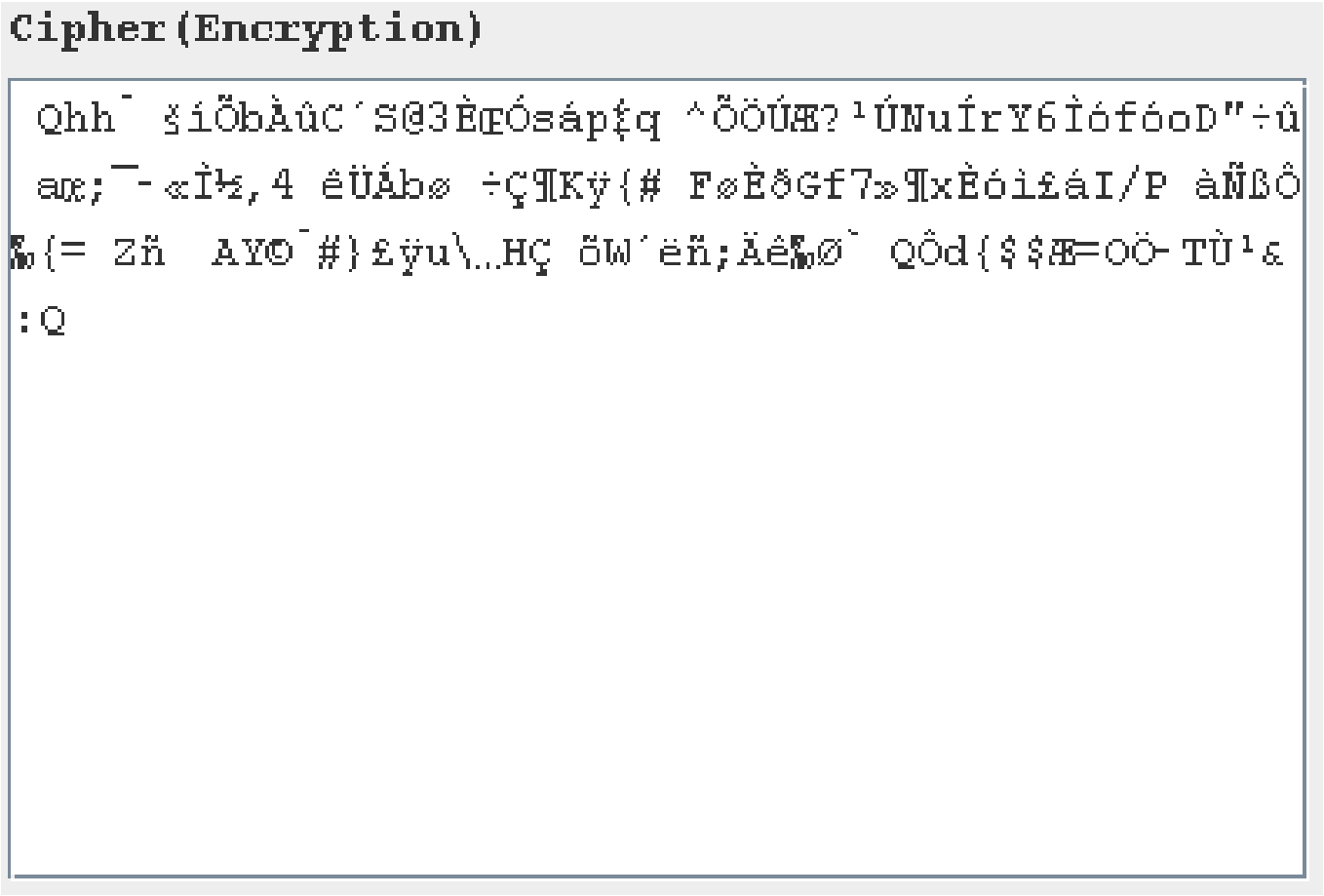

**NOTE:** in cipher above there are some characters disappear.

Now, suppose the receiver wants to do decryption to knows the original message M !, our decryption approach illustrates the mechanisem as the following:

1- The binary representation of the cipher C (character by character) is the same as in step (4)above in example(every representation of character must be byte): L=
2- 1000000001010001011010000110100010001000000000001100110101110110111010101
0000110101100010110000000111110110100001110110100010100110100000000110011
1100100010001100110100110111001111100001011100001000011101110001000001 01
0101111011010101110101101101101011000110001111111011100111011010010011 10
0111010111001101011100100101100100110110110011001111001101100110111100 11
0110111101000100001000101111011111111011000111110110000110011100001110 11
1010111110010110101010111100110010111101001011000011010010010000111010 10
1101110011000001011000101111100000010001111101111100011110110110010010 11
1111111011110110010001100000111010001101111100011001000111100000100011 1
0110011000110111101110111011011001111000110010001111001111101100101000 11
1110000101001001001011110101000001111111111000001101000111011111110101 00
1000100101111011001111010000000100101101011110001000011100000001010100 0001

0101100110101001100010000010001101111101101000111111111011101010101110010000101010010001100011110001101111101010101011110110100111010111111000100111011110001001110101010001001110110001001100000011111010100011101010001100100011110110010010000100100110001100011110101001111110101101001011001010100110110011011100100100110001110100101000 1..

3- The dicimal representation of L is $M_2$= 86306282674243970028653533390575702972020736691095525172358216933099888933418710146515279277996702436706352300213627549860704339714293985064495785166656115475126057162348126489314727983300016280264694240023381924156653359707066914540655040456516989913345307746045840447110479114302210672152560261189308933148968553121765906428348893974176562446953651026194745 13.

4- Applying RSA Decryption on $M_2$ and get $M_1$ the same as in the step (2) in encryption algorithm above as following:
$M_1 = M_2^d (mod\ n) =$
674779901925981090940264347938001282716237591256227670169614914057960882743836443699701318079498138580225881765380967217303146583780732812656111452862620735498353002748845442215768238953985343249300352185989716276564725635283466626213032688391167351175804072921362768408803847685904489446625178407924866937386332916 39357.

5- Convert $M_1$ into binary representation(adding zeros to the left of representation to be the length divided by 8) and we will get S the same in the first step in encryption algorithm above as following: S=
0101011101100101001000000111001001100101011000110110111101101101011011010110010101101110011001000000100000011101010111001101101001011011100110011100100000001110100011010000110100101110011001000000110000101110000001110000011100100110111101100001011000110110100000010000000111010001101111001000000111001101100101011000110111010101110010011001010010000001110100011010000110010100100000011010010110110101110000011011110111001001110100011000010110111001110100001000000110100101101110011001100110111101110010011011010110000101110100011010010110111101101110001000000111011101101000011010010110001101101000001000000110010101111000011000110110100001100001011011100110011101100101011001000000100000011011110111011001100101011100100010000001100101001011010110110101100001011010010110110001110011001000000110000101101110011001000000100000011010010110111001110100011001010111001001101110011001010111010000100000011011100110010101110100010001110111011011110111001001101011011100110010111000100001010000000001100010010011001011110001100110101111110010101100101111010110110010100000101010001111 01.

6- Divide the last representation into bytes and you able to do that because the number of digits is divided by 8.

7- Convert from left to right every byte in the last representation into associated character in ASCII and will be get the exactly message "M" :

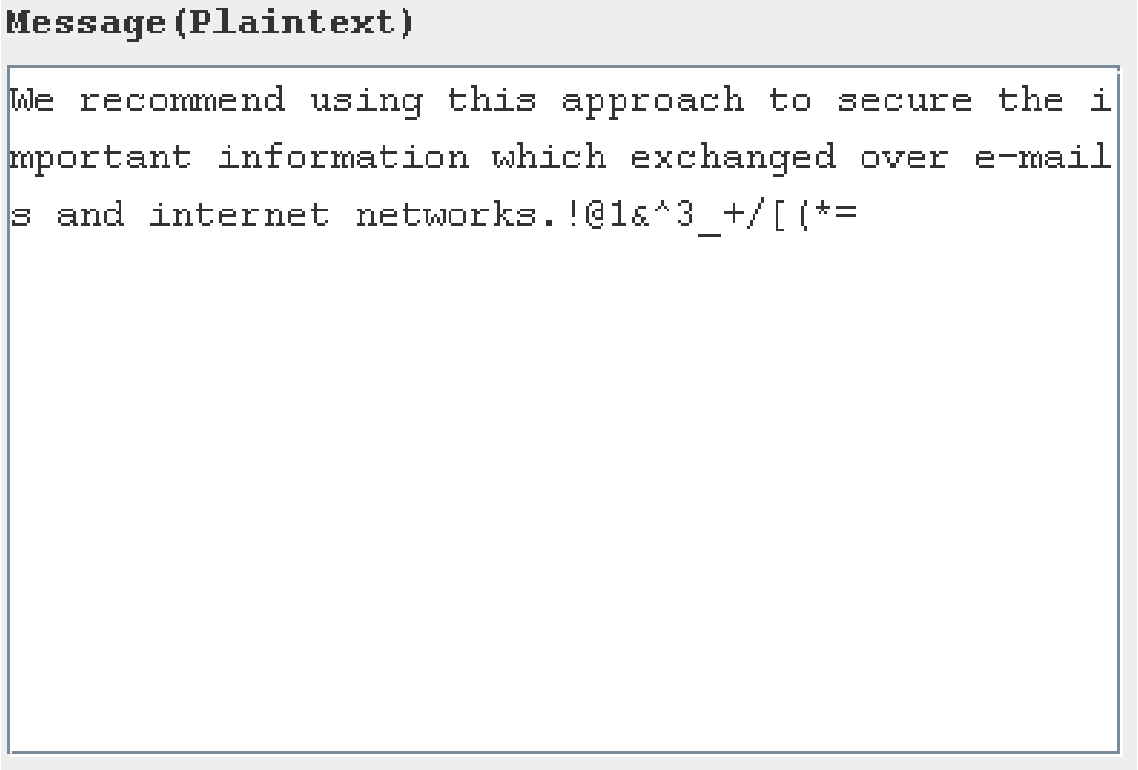

**NOTE:** We can see that decryption approach is absolutely true and will lead to the exactly original message, because it was built from last step to first step in the proposed encryption approach in this paper(it is the inverse of encryption approach step by step!).

## 6. Java Application For Proposed Approach

Depending on java language[4][5] we build an application to implimination our approach. In this application we can generate d1,d2-bits primes numbers p and q depending on Miller–Rabin Test algorithm[4][2], and we can directly apply RSA algorithm with our approach to encrypt and decrypt message text according to ASCII. Also, this application can tell us directly if the RSA algorithm is applicable with this approach or not(just when we put the message text wich we want incrypt it).

Figure(1) reviews the results of the numerical example mentioned above by our application, and the figure(2) appeares when the "go to decrypt the message" button is preesed to come back the original message according to the private key "d":

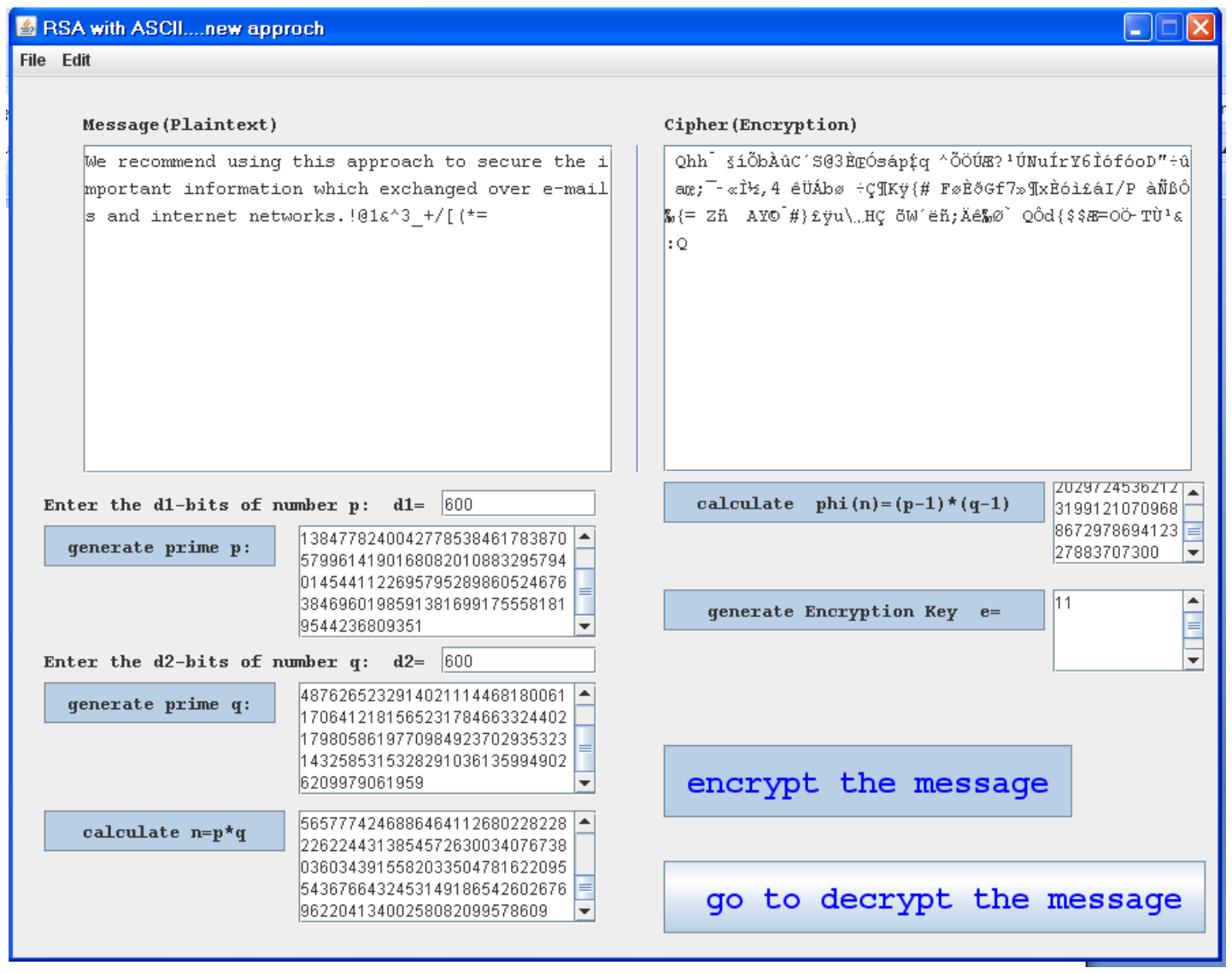

Figure(1): java application for encryption approach deppending on RSA and ASCII

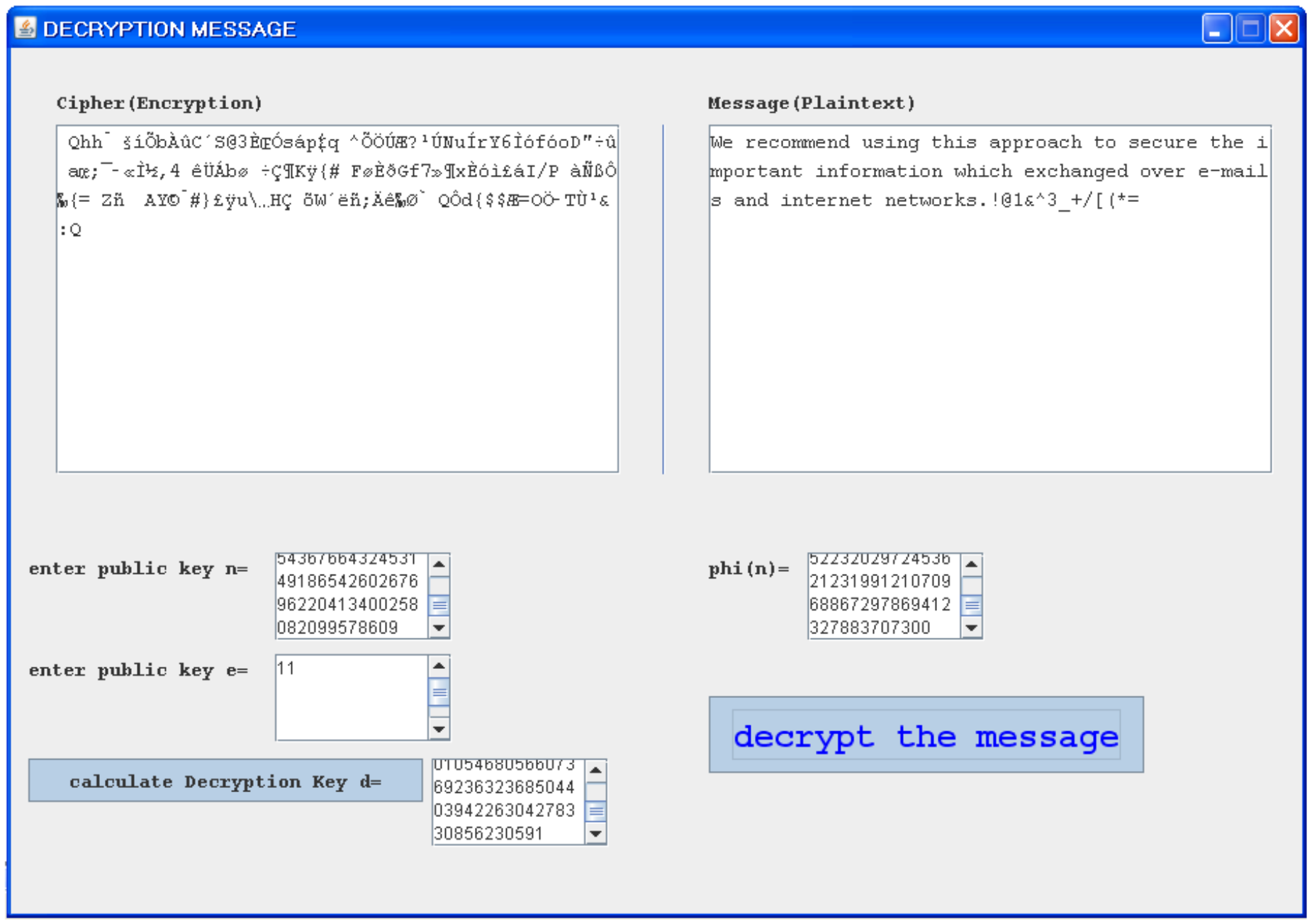

Figure(2): java application for decryption approach deppending on RSA and ASCII

## 7. Conclusion And Future Plan

This paper describes a new approach for encryption and decryption message text using RSA algorithm depending on ASCII, and java application to implementation this approach. We recommend using this approach to secure the important information which exchanged over e-mails and internet networks, even, secure the gibberish passwords.

Since the RSA algorithm will be slow when the message text is big, so, we recommend to use this approach after dividing the message into blocks and apply this approach on the every block in fast way while transmitting in the networks ...block by block, so we can build java applet to do that and this is will be our future plan with what the relationship between the modulo "n" in RSA and the number representation of the message to allow us dividing into blocks automatically in an efficient way.

## AUTHORS

**Ahmad Steef** is a PhD Candidate in mathematics at Al-Baath University, Homs, Syria.

**Mohammad Nour Shamma** is a professor in mathematics at Damascus University- Basic Sciences department- the Mechanical and Electrical Engineering- Syria. His Research interests include mathematical analysis, Cryptography.

**Abdulbaset Alkhatib** is a professor in mathematics at Al-Baath University- Homs, Syria. His Research interests include Linear algebra, Cryptography.